\documentclass{pasj00}  
\begin{document}
\SetRunningHead{M.Shimizu et al.}{Reliability of Merger Trees in the  Monte-Carlo Modeling of Galaxy Formation}
\Received{2001 November 29}%{yyyy/mm/dd}
\Accepted{2002 July 11}%{yyyy/mm/dd}

\title{Reliability of Merger Tree Realizations of Dark Halos \\
  in the  Monte-Carlo Modeling of Galaxy Formation}

\author{Mamoru \textsc{Shimizu},\altaffilmark{1} Tetsu
  \textsc{Kitayama},\altaffilmark{2} Shin
  \textsc{Sasaki},\altaffilmark{3} and Yasushi {\sc
    Suto}\altaffilmark{1,4}}  \altaffiltext{1}{Department of Physics,
  School of Science, The University of Tokyo, Tokyo 113-0033}
\email{mshimizu@utap.phys.s.u-tokyo.ac.jp} \altaffiltext{2}{Department
  of Physics, Toho University,  Funabashi, Chiba 274-8510}
\altaffiltext{3}{Department of Physics, Tokyo Metropolitan University,
  Hachioji, Tokyo 192-0397}  \altaffiltext{4}{Research Center for the
  Early Universe, School of Science, The University of Tokyo, Tokyo
  113-0033}


\maketitle

\begin{abstract}
  We examine the reliability of the merger trees generated for the
  Monte-Carlo modeling of galaxy formation. In particular we focus on
  the cold gas fraction predicted from merger trees with different
  assumptions on the progenitor distribution function, the timestep,
  and the mass resolution.  We show that the cold gas fraction is
  sensitive to the accuracy of the merger trees at small-mass scales
  of progenitors at high redshifts. One can reproduce the
  Press--Schechter prediction to a reasonable degree by adopting a
  fairly large number of redshift bins, $N_{\rm step}\sim 1000$, in
  generating merger trees, which is a factor of ten larger than the
  canonical value used in previous literature.
\end{abstract}

\section{Introduction}

Understanding the formation and evolution of galaxies is a fundamental
step in linking the initial condition of the universe and the
cosmological observational data.  Recent systematic studies of
high-redshift objects, such as quasars and Lyman-break galaxies, should
provide important clues to the early universe, although their proper
interpretation is often not so straightforward, mainly because those
objects certainly do evolve in time.

A theoretical study of galaxy evolution, especially its spectroscopic
evolution, from a cosmological context, was begun by \citet{tinsley80} and
followed by many authors (e.g., \cite{Bruzual83}; \cite{AY86};
\cite{GR87}; \cite{CB91}; \cite{BC93}; \cite{KA97}). These studies are
based on a so-called `one-zone' model which assumes that a galaxy does
not interact with other galaxies. It is now fairly established,
however, that structures in the universe have built up hierarchically
from small to large scales as in a cold dark matter (CDM) model. This
means that a galaxy interacts and sometimes merges with other galaxies
even if it was an isolated system at birth. The predictions in the
one-zone model therefore may be significantly different from what
happened to galaxies in a hierarchical universe.

White and Frenk (1991) developed a detailed analytic formalism to
describe the formation and evolution of galaxies while taking account of the
hierarchical merging of dark-matter halos, gas cooling, star formation,
and supernova feedback.  Subsequent numerical approaches in modeling
hierarchical merging of dark halos employ two somewhat different
algorithms; one is called the `block model' in which a random-Gaussian
density fluctuation field is generated by dividing a hypothetical
rectangular box recursively (\cite{CK88}; \cite{Cole91};
\cite{Cole94}). While this algorithm is simple and straightforward,
the resulting halo masses are necessarily binned in discrete steps of
a factor of two. The other generates a realization of halo merger
trees according to a probability distribution function predicted by
the extended Press--Schechter theory (\cite{Bower91}; \cite{Bond91};
\cite{KW93}; \cite{SK99}; \cite{SL99}).  The latter is widely used in
studying the cosmological evolution of galaxies in a hierarchical
universe (\cite{KWG93}; \cite{baugh98}; \cite{SP99}; \cite{Cole00};
\cite{nagashima}). Throughout the present paper, we call the latter
method the Monte-Carlo modeling of merger histories (simply, the
Monte-Carlo modeling), while it is usually referred to as a
semi-analytic model of galaxy formation (SAM).

The most important ingredient in Monte-Carlo modeling is the
conditional joint-probability distribution function of a set of
\textit{progenitor} halos of mass $M_2^{j}$ at a redshift of $z_2$, which
is a part of a \textit{parent} halo of mass $M_1$ at $z_1$,
conceptually written as
%%%%%%%%%%%%%%%%%%%%%%%%%%%%%%%%%%%%%%%%%%%%%%%%%%%%%%%%%%%%%%%%%%%%%%%%%
\begin{eqnarray}
\label{eq:jointprob}
{\rm Prob}(M_2^1, M_2^2, \cdots, M_2^N, z_2 |
M_1, z_1) dM_2^1 dM_2^2 \cdots dM_2^N \cr
\qquad (N=1, \cdots , \infty).  
\end{eqnarray} 
%%%%%%%%%%%%%%%%%%%%%%%%%%%%%%%%%%%%%%%%%%%%%%%%%%%%%%%%%%%%%%%%%%%%%%%%%
Unfortunately only an analytical expression for the conditional
one-point probability distribution function, Prob($M_2^i$, $z_2 |M_1$,
$z_1$), is known based on the extended Press--Schechter theory
(for the special case of the Poisson initial power spectra, see a
different approach by \cite{SL99}); one thus needs to employ an
additional \textit{assumption} in generating realizations of merger
trees of halos in general (e.g., \cite{KW93}; \cite{SK99}).
Furthermore, any numerical procedure to generate them necessarily
involves several \textit{ad hoc} parameters due to the limitation of
the available computation resources including the finite timestep of
computation, the minimum mass of halos to be included in merger trees,
and the maximum number of progenitors for each halo at each step.

The purpose of the paper is to perform a systematic investigation of
possible artificial effects of the above-mentioned problems on merger tree
realizations, and to re-examine the validity of the Monte-Carlo
modeling. In particular, we focus on the extent to which the resulting
merger trees reproduce the conditional one-point probability
distribution function predicted by the extended Press--Schechter theory,
which directly changes the fraction of cold gas.  Exactly for this
reason, we adopt a conventional $\Lambda$CDM model with the
cosmological parameters $\Omega_{0}=0.3$, $\lambda_{0}=0.7$, $h=0.7$,
$\sigma_{8}=1.0$, and $\Omega_{\mathrm{B}}=0.015h^{-2}$ (e.g.,
Kitayama, Suto 1997; Kitayama et al.\ 1998), and neglect star
formation and a feedback effect for definiteness.

\section{Merger Trees of Dark Matter Halos}

\subsection{Constructing Merger Trees of Dark-Matter Halos
\label{subsec:construction}}

Our model of merging histories of dark-matter halos is mainly based on
that of Somerville and Kolatt~(1999), which we adopt as our fiducial choice and
slightly modify their original scheme as follows. We begin with a halo
of mass of $M_1= M_{\mathrm{root}}$ at a redshift of $z_1=z_{\mathrm{min}}$,
and consider its progenitors at a slightly earlier redshift of 
$z_2=z_1+\Delta z(z_1)$.  Since the joint conditional probability for
the progenitors [equation~(\ref{eq:jointprob})] is not known, we choose the
$i$-th progenitor halo of mass $M_2^i$ according to the
\textit{one-point} conditional probability, Prob($M_2^i$, $z_2 |M_1$,
$z_1$), as long as $M_2^i > M_{\rm res}$ and the total mass satisfies
%%%%%%%%%%%%%%%%%%%%%%%%%%%%%%%%%%%%%%%%%%%%%%%%%%%%%%%%%%%%%%%%%%
\begin{eqnarray}
\label{eq:massconserve}
\sum_{i=1}^N M_2^i < M_1 - \Delta M_{\rm acc}(<M_{\rm res}) ,
\end{eqnarray}
%%%%%%%%%%%%%%%%%%%%%%%%%%%%%%%%%%%%%%%%%%%%%%%%%%%%%%%%%%%%%%%%%%
where
%%%%%%%%%%%%%%%%%%%%%%%%%%%%%%%%%%%%%%%%%%%%%%%%%%%%%%%%%%%%%%%%%%
\begin{eqnarray}
\Delta M_{\rm acc}(<M_{\rm res})
= \int_{0}^{M_{\rm res}}\!\!\!\!dM_2 M_2 \frac{dN}{dM_2}(M_2,z_2|M_1,z_1)
\end{eqnarray}
%%%%%%%%%%%%%%%%%%%%%%%%%%%%%%%%%%%%%%%%%%%%%%%%%%%%%%%%%%%%%%%%%%
is the expectation value of the total mass of halos smaller than the
resolution mass ($M_{\rm res}$) with $dN/dM_2(M_2,z_2|M_1,z_1)$ being
the appropriate conditional mass function [equation~(\ref{eq:eps-num})
below]. In other words, we distinguish the discrete merging and the
continual accretion at mass $M_{\rm res}$, and do not resolve the
halos below $M_{\rm res}$ in our merger trees.  Once all relevant
progenitor halos are selected, we repeat the above procedure
recursively for each progenitor until the maximum redshift ($z_{\rm
  max}$).  Unless otherwise stated, we set $z_{\mathrm{min}}=0$ and
$z_{\mathrm{max}}=15$ in the present paper. For convenience, we list
in table~\ref{tab:parameters} variables which are extensively
discussed in the present paper.

In the original method by Somerville and Kolatt~(1999), one stops selecting
progenitors when $M_1 - \sum_{i=1}^N M_2^i$ becomes less than $M_{\rm
  res}$, but without imposing the condition $M_2^i>M_{\rm res}$. They
carefully tuned the timesteps depending on $M_{1}$ so that the
resulting progenitor mass function becomes close to equation
(\ref{eq:eps-num}) below.  Rather, we stop choosing the progenitor when
$M_1 - \Delta M_{\rm acc}(<M_{\rm res}) - \sum_{i=1}^N M_2^i$ becomes
negative, and the last selected progenitor $M_2^N$ is not included in
the tree. In this case, the remaining mass $M_1 - \Delta M_{\rm
  acc}(<M_{\rm res}) - \sum_{i=1}^{N-1} M_2^i$ is not necessarily
smaller than $M_{\rm res}$.  We find that our method reproduces
equation (\ref{eq:eps-num}) even with the $M_1$-independent timesteps.

\subsection{Conditional Probability Distribution Function}

The most important and subtle issue is the proper choice of the
\textit{one-point} conditional probability, Prob($M_2$, $z_2 |M_1$,
$z_1$).  Bower (1991) and Bond et al.\ (1991) derived the conditional
probability of $M_2$ at $z_2$, which is a part of halo $M_1$ at $z_1$:
%%%%%%%%%%%%%%%%%%%%%%%%%%%%%%%%%%%%%%%%%%%%%%%%%%%%%%%%%%%%%%%%%%%%%%
\begin{eqnarray}
\label{eq:eps-mass}
 \frac{dP}{dM_{2}}(M_{2},z_{2}|M_{1},z_{1})
 &=& 
\frac{\delta_{\rm c,2}-\delta_{\rm c,1}}{\sqrt{2\pi(S_{2}-S_{1})^{3}}}\
\cr
 && \exp\!\left[
 -\frac{(\delta_{\rm c,2}-\delta_{\rm c,1})^{2}}{2(S_{2}-S_{1})}
 \right]
 \left|
 \frac{dS_{2}}{dM_{2}}
 \right| ,
\end{eqnarray}
%%%%%%%%%%%%%%%%%%%%%%%%%%%%%%%%%%%%%%%%%%%%%%%%%%%%%%%%%%%%%%%%%%%%%%
where $\delta_{\mathrm{c},i} \sim 3 (12\pi)^{2/3} /20 D(z_i)$ (its
useful approximate formula may be found in Kitayama, Suto 1996) is the
critical over-density of the mass density field at a redshift of $z_{i}$,
$D(z_i)$ is the linear growth rate, and $S_{i}\equiv\sigma^{2}(M_{i})$ is
a mass variance of the density field top-hat smoothed over the mass scale
$M_i$.  Since equation (\ref{eq:eps-mass}) is the
\textit{mass-weighted} probability for $M_2$, it is easily translated
to the \textit{number-weighted} probability that we need in the halo
number counting:
%%%%%%%%%%%%%%%%%%%%%%%%%%%%%%%%%%%%%%%%%%%%%%%%%%%%%%%%%%%%%%%%%%%%%%
\begin{equation}
 \label{eq:eps-num}
 \frac{dN}{dM_{2}}(M_{2},z_{2}|M_{1},z_{1})
=  \frac{M_{1}}{M_{2}}\frac{dP}{dM_{2}}(M_{2},z_{2}|M_{1},z_{1}) .
\end{equation}
%%%%%%%%%%%%%%%%%%%%%%%%%%%%%%%%%%%%%%%%%%%%%%%%%%%%%%%%%%%%%%%%%%%%%%

%%%%%%%%%%%%%%%%%%%%%%%%%%%%%%%%%%%%%%%%%%%%%%%%%%%%%%%%%%%%%%%%%%%
\begin{figure}[thb]
  \centering \FigureFile(80mm,80mm){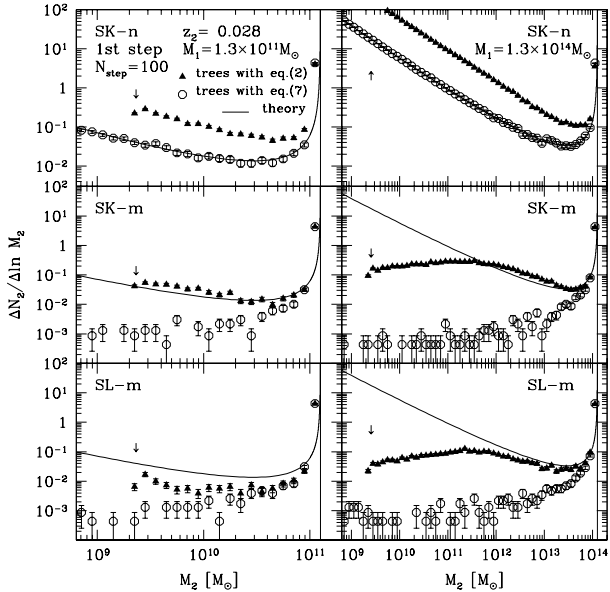}
\caption{Progenitor number distribution at the 1st timestep
[$z_{\scriptscriptstyle(1)}=0.028$ with $N_{\rm step}=100$ in
equation~(\ref{eq:zbin})] for $M_{\mathrm{root}}=1.3\times 10^{11}\;M_{\odot}$
(\textit{Left}) and $1.3\times 10^{14}\;M_{\odot}$ (\textit{Right}).  Top
(SK-n) and middle (SK-m) panels adopt the number-weighted and
mass-weighted conditional progenitor probability functions
[equations~(\ref{eq:eps-num}) and (\ref{eq:eps-mass})] according to
Somerville and Kolatt~(1999), while bottom (SL-m) panels use the mass-weighted one
according to Sheth and Lemson~(1999).  The theoretical prediction in the extended
Press--Schechter theory [equation~(\ref{eq:eps-num})] is plotted in solid
curves. The solid triangles and open circles indicate the averages over
$N_{\rm ens}=10^4$ merging tree realizations with the different mass
conservation prescriptions, equations~(\ref{eq:massconserve}) and
(\ref{eq:nmassconserve}), respectively. The quoted error bars represent
the Poisson error in each mass bin, and the arrows indicate the value of
$M_{\rm res}$.  \label{fig:OneStep}}
\end{figure}
%%%%%%%%%%%%%%%%%%%%%%%%%%%%%%%%%%%%%%%%%%%%%%%%%%%%%%%%%%%%%%%%%%%

\begin{figure}[thb]
  \centering \FigureFile(80mm,80mm){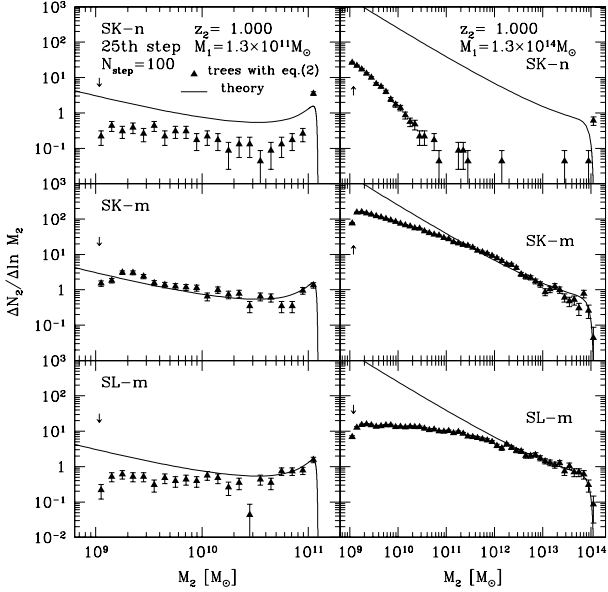} 
\caption{Same as figure~\ref{fig:OneStep} except at the 25th timestep
($z_{\scriptscriptstyle(25)}=0.028$) for the mass conservation
prescription [equation~(\ref{eq:massconserve})]; $M_{\mathrm{root}}=1.3\times
10^{11}\;M_{\odot}$ (\textit{Left}) and $1.3\times 10^{14}\;M_{\odot}$
(\textit{Right}).  \label{fig:100Step}}
\end{figure}

Figures~\ref{fig:OneStep} and \ref{fig:100Step} present how the
progenitor number distribution of the merger tree realizations
reproduces the theoretical prediction: $M_{\mathrm{root}}=1.3\times
10^{11}\;M_{\odot}$ (\textit{Left}) and $1.3\times 10^{14}\;M_{\odot}$
(\textit{Right}).  In these plots, we adopt the logarithmically equal
timestep in redshift:
%%%%%%%%%%%%%%%%%%%%%%%%%%%%%%%%%%%%%%%%%%%%%%%%%%%%%%%%%%%%%%%%%%%%%
\begin{eqnarray}
\label{eq:zbin}
  z_{\scriptscriptstyle(i)}=(1+z_{\mathrm{min}})\times
  \left(
    \frac{1+z_{\mathrm{max}}}{1+z_{\mathrm{min}}}
  \right)^{i/N_{\mathrm{step}}}-1  \cr
\qquad (i=1, \cdots , N_{\rm step}),
\end{eqnarray}
%%%%%%%%%%%%%%%%%%%%%%%%%%%%%%%%%%%%%%%%%%%%%%%%%%%%%%%%%%%%%%%%%%%%%
where $N_{\mathrm{step}}$ is the total number of the redshift bins.
We defer the discussion concerning the choice of $N_{\mathrm{step}}$ to the
next subsection (\ref{sec:timestep}), and fix
$N_{\mathrm{step}}=100$ throughout this subsection.

The top panels in figure~\ref{fig:OneStep} show the result for the 1st
timestep ($i=1$) corresponding to $z_{\scriptscriptstyle(1)}=0.028$
according to equation (\ref{eq:zbin}).  We call this model SK-n
indicating the Somerville and Kolatt~(1999) method with the number-weighted
probability. The symbols indicate the average $(M_2/N_{\rm
  ens})(\Delta N_2/\Delta M_2)$ with the quoted error-bars being the
corresponding one-sigma dispersion, where $\Delta N_2$ is the number
of progenitors in the range of mass $M_2 \sim M_2 + \Delta M_2$, and
we adopt $\Delta \log_{10}M_2 = 0.1$.  In the SK-n model we generate
the random numbers according to the number-weighted probability
distribution function [equation~(\ref{eq:eps-num})]. Also we have to set the
lower limit on the progenitor mass in adopting the SK-n model so as to
avoid a divergent total probability. We adopted the lower limit of
$10^{-3}M_{\rm res}$,  and made sure that the mass function of
progenitors of the mass range of our interest $M>M_{\mathrm{res}}$ is
properly reproduced.

The solid triangles show our result based on the algorithm outlined in
the previous subsection. Somewhat surprisingly, they are completely
different from the theoretical distribution that we use in generating
the trees (solid curve). Note that figure~\ref{fig:OneStep} plots the
number distribution multiplied by $M_2$, $M_2 \, dN/dM_2= M_1 \,
dP/dM_2$ [see equation~(\ref{eq:eps-num})]. To understand the origin of the
discrepancy, we generate the progenitors at the 1st timestep for
$N_{\mathrm{ens}}$ realizations simultaneously as long as they satisfy
%%%%%%%%%%%%%%%%%%%%%%%%%%%%%%%%%%%%%%%%%%%%%%%%%%%%%%%%%%%%%%%%%%
\begin{eqnarray}
\label{eq:nmassconserve}
\sum_{i=1}^{N'} M_2^i < N_{\mathrm{ens}} 
\left[M_1 - \Delta M_{\rm acc}(<M_{\rm res}) \right] ,
\end{eqnarray}
%%%%%%%%%%%%%%%%%%%%%%%%%%%%%%%%%%%%%%%%%%%%%%%%%%%%%%%%%%%%%%%%%%
instead of the mass conservation [equation~(\ref{eq:massconserve})] for each
individual parent halo. In the above, $N'$ is not the number of
progenitors for a single halo at $z=z_{\rm min}$, but for an ensemble
of $N_{\rm ens}$ halos with the same mass $M_{\rm root}$.  The
resulting distribution is plotted in open circles, and in fact shows
good agreement with the theoretical curve.

This is simply because we attempt to generate a joint distribution of
progenitors with a repeated use of the conditional probability
[equation~(\ref{eq:eps-num})] \textit{incorrectly}; except for the first
progenitor, the mass conservation for each halo
[equation~(\ref{eq:massconserve})] introduces an additional cutoff at higher
mass in the selection probability of progenitors.  In fact the
conditional probability [equation~(\ref{eq:eps-num})] for $(z_2-z_1)/z_1 \ll
1$ is sharply peaked at a mass scale $M_2$ just below the parent mass
$M_1$, and thus even a small value of the first progenitor mass may
effectively bias not to choose remaining progenitors in the peak.
Thus, the resulting distribution is significantly biased toward
low-mass objects, i.e., the number density of the low-mass objects
exceeds 
the theoretical predictions by an order of magnitude (top panels in
figure~\ref{fig:OneStep}).

One way out of this problem is to generate many ($>100$) realizations
simultaneously, as Kauffmann and White~(1993) adopted. Even in this case, one needs
to specify an additional assumption on how to plant a set of
progenitors in a single merger tree \textit{by hand}.  Moreover, a
practical implementation of this method requires one to discretize the
halo mass, and thus becomes computationally demanding as  both the mass
and time resolutions increase.

Another possibility is to artificially distort the input conditional
probability so that the selected progenitors obey the distribution
[equation~(\ref{eq:eps-num})]. While the required correction may be a fairly
definite mathematical problem, we do not know the exact answer, and
thus have to proceed in a phenomenological fashion. Basically this is
the approach taken by Somerville and Kolatt~(1999)  and Sheth and
Lemson~(1999), who adopted the 
\textit{mass-weighted} probability [equation~(\ref{eq:eps-mass})] as the
theoretical input.  The middle and bottom panels in
figure~\ref{fig:OneStep} show the resulting distribution for SK-m
(\cite{SK99}-mass weighted) and SL-m (\cite{SL99}-mass weighted),
respectively. Clearly the resulting distributions (filled triangles)
become much closer to equation (\ref{eq:eps-num}) under the constraint
[equation~(\ref{eq:massconserve})] although their original distributions
[i.e., without the constraint (\ref{eq:massconserve})] plotted in open
circles are completely different.

As this indicates, the input conditional probability for the current
purpose should be small at lower mass scales of $M_2$ relative to
equation (\ref{eq:eps-num}). Thus, we also attempted to make the probability
proportional to $(M_1/M_2)^\alpha (dP/dM_2)$. Note that the mass- and
number-weighted probabilities [equations~(\ref{eq:eps-mass}) and
(\ref{eq:eps-num})] correspond to $\alpha=0$ and $1$. We were not able
to obtain a similar degree of agreement for a value of $\alpha$ very
different from 0, but did not find a significant change for
$-0.2\lesssim \alpha \lesssim +0.2$. Thus, we decided to adopt
$\alpha=0$ (the mass-weighted probability) as Somerville and Kolatt~(1999). This
choice has an advantage that a numerical routine to generate random
numbers becomes easier than the cases of $\alpha \not= 0$. Define $x_2
= (\delta_{\rm c,2}-\delta_{\rm c,1})/\sqrt{S_{2}-S_{1}}$ that obeys
Gaussian distribution of a unit variance. Equation (\ref{eq:eps-mass})
implies that the desired distribution of $M_2$ can be simply given via
$S_2 = S(M_2) = S_1 + (\delta_{\rm c,2}-\delta_{\rm c,1})^2/x_2^2$.
Since Gaussian-distributed random numbers can be implemented
easily, all we have to do is to supply the inverse function of the
mass variance, $M_2 = S^{-1}(S_2)$. Our SK-m implementation seems to
yield a larger discrepancy between theory and the merger tree
realizations at small $M_2$ regimes than their original results. This
may be due to the different choice of the timestep and the condition
how to stop selecting progenitor halos. In any case, however, this
discrepancy rapidly fades away in constructing the merger tree using
many timesteps, as we show in figures~\ref{fig:100Step} and
\ref{fig:1000Step}.

%%%%%%%%%%%%%%%%%%%%%%%%%%%%%%%%%%%%%%%%%%%%%%%%%%%%%%%%%%%%%%%%%%%%%%
%%%%%%%%%%%%%%%%%%%%%%%%%%%%%%%%%%%%%%%%%%%%%%%%%%%%%%%%%%%%%%%%%%%%%%

Figure~\ref{fig:100Step} plots a snapshot of the progenitor
distribution at the 25th timestep ($z=z_{\scriptscriptstyle(25)}=1.0$),
which makes sure that the mass-weighted probability reasonably works,
even when we trace the merger tree by many steps. Also SK-m works a bit
better than SL-m especially at small mass scales.  Originally SL-m was
proposed to correct for the halo exclusion effect, but does not work so
efficiently at least in the range of parameters we surveyed.
Figure~\ref{fig:100Step} also indicates that SK-n is substantially
different from the analytical solution.  This is due to the fact that
SK-n tends to select relatively less massive progenitors preferentially
(see figure~\ref{fig:OneStep}), and this tendency simply accumulates in
many steps. On the contrary, the behavior of SK-m and SL-m becomes
closer to the analytical solution than in the case of
figure~\ref{fig:OneStep}. This is because the latter two models
well 
approximate the probability distribution around $M_1$, which is the most
important range when constructing real merger trees with many
timesteps.

\subsection{Timestep \label{sec:timestep}}

The next question that we address is the appropriate choice of the
timestep. While this is an equally important problem in the
Monte-Carlo modeling, the previous authors did not discuss it in an
explicit manner.

%%%%%%%%%%%%%%%%%%%%%%%%%%%%%%%%%%%%%%%%%%%%%%%%%%%%%%%%%%%%%%%%%%%%%
\begin{figure}[thb]
  \centering\FigureFile(60mm,60mm){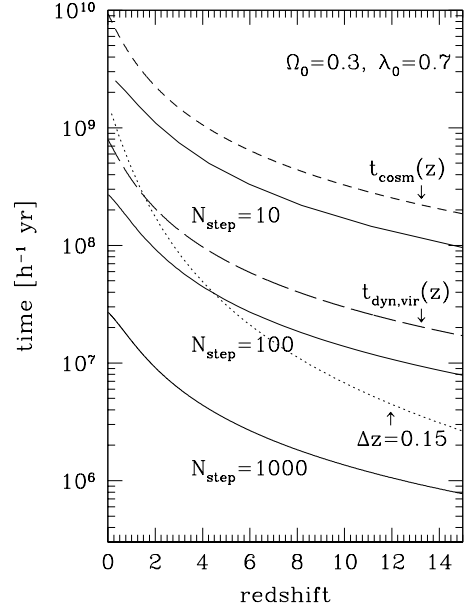}
  \caption{Comparison among various timescales as a function of $z$ in
    $\Lambda$CDM.  The short-dashed and long-dashed curves indicate the
    cosmic time at $z$, and the dynamical time of dark halos just
    virialized at $z$.  The dotted curve corresponds to the timesteps
    in linearly equal intervals in $z$, i.e., 100 bins between $z_{\rm
      min}=0$ and $z_{\rm max}=15$.  The three solid curves correspond to
    the timesteps in logarithmically equal intervals in $z$; $N_{\rm
      step} =10$, 100, and 1000 from top to bottom.
  \label{fig:timescale}}
\end{figure}
%%%%%%%%%%%%%%%%%%%%%%%%%%%%%%%%%%%%%%%%%%%%%%%%%%%%%%%%%%%%%%%%%%%%%

Obviously, the timestep needs to be smaller than the dynamical
timescale of halos just virialized at the redshift, $t_{\rm dyn,
  vir}(z)$, because they are the objects that serve as the initial
condition for the Monte-Carlo modeling.  Figure~\ref{fig:timescale}
compares this timescale and our choice [equation~(\ref{eq:zbin})] for
$N_{\rm step}=10$, 100, 1000. Also, we plot the cosmic time $t_{\rm
  cosm}(z)$ and the timestep corresponding to the linearly equal bin
($\Delta z=0.15$).  Even this simple comparison indicates that the
logarithmic time bin with $N_{\rm step} \gtrsim 100$ is required.

While the important question is how small timescales one should
resolve, it critically depends on the problem that one would like to
address. Therefore, we rather ask how many timesteps we need to
reproduce the progenitor distribution.

%%%%%%%%%%%%%%%%%%%%%%%%%%%%%%%%%%%%%%%%%%%%%%%%%%%%%%%%%%%%%%%%%%%%%%%%%
\begin{figure}[thb]
  \centering \FigureFile(80mm,80mm){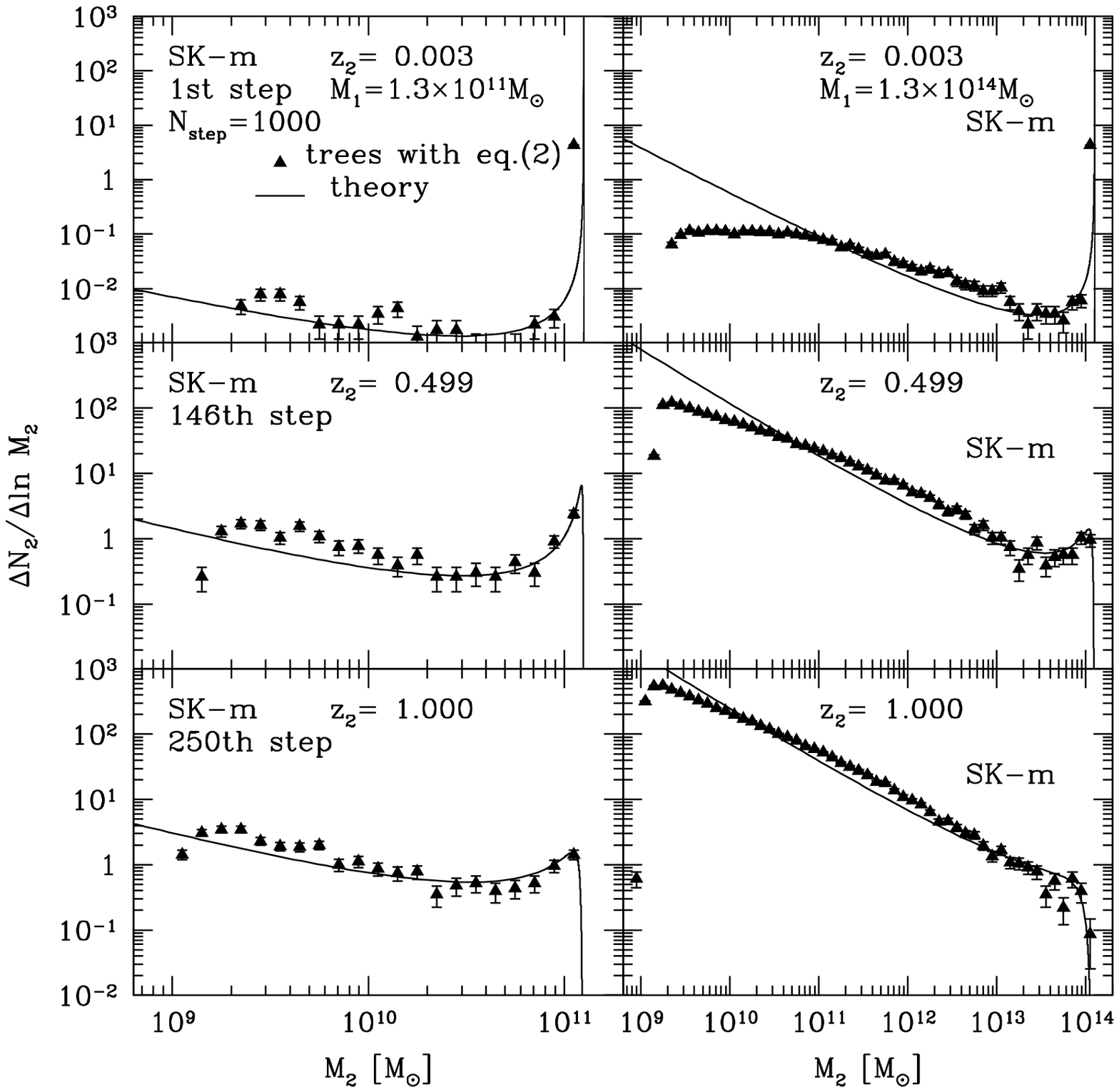}
  \caption{Progenitor number distribution functions at the 1st, 146th
    and  250th steps from the $N_{\rm step}=1000$ merger trees using
    the SK-m method; $z_{\scriptscriptstyle(1)}=0.003$
    (\textit{Upper}), $z_{\scriptscriptstyle(146)}=0.499$
    (\textit{Middle}), and $z_{\scriptscriptstyle(250)}=1.0$
    (\textit{Lower}).  The left and right panels are for
    $M_{\mathrm{root}}=1.3\times 10^{11}\;M_{\odot}$ and $1.3\times
    10^{14}\;M_{\odot}$.  The theoretical prediction in the extended
    Press--Schechter theory [equation~(\ref{eq:eps-num})] is plotted as solid
    curves.
    \label{fig:1000Step}}
\end{figure}
%%%%%%%%%%%%%%%%%%%%%%%%%%%%%%%%%%%%%%%%%%%%%%%%%%%%%%%%%%%%%%%%%%%%%%%%%

To see explicitly how different values of $N_{\rm step}$ affect the
realizations of merger trees, we plot in figure~\ref{fig:1000Step} the
progenitor distribution with $N_{\rm step}=1000$ at redshifts of 
$z_{\scriptscriptstyle(1)}=0.003$,
$z_{\scriptscriptstyle(146)}=0.499$, and
$z_{\scriptscriptstyle(250)}=1.0$. A comparison of
figures~\ref{fig:100Step} and \ref{fig:1000Step} indicates that the
average progenitor distribution is indeed slightly better reproduced
by $N_{\rm step}=1000$ than $N_{\rm step}=100$, particularly at small
mass scales.

%%%%%%%%%%%%%%%%%%%%%%%%%%%%%%%%%%%%%%%%%%%%%%%%%%%%%%%%%%%%%%%%%%%%%%%%%
\begin{figure}[thb]
  \centering \FigureFile(80mm,80mm){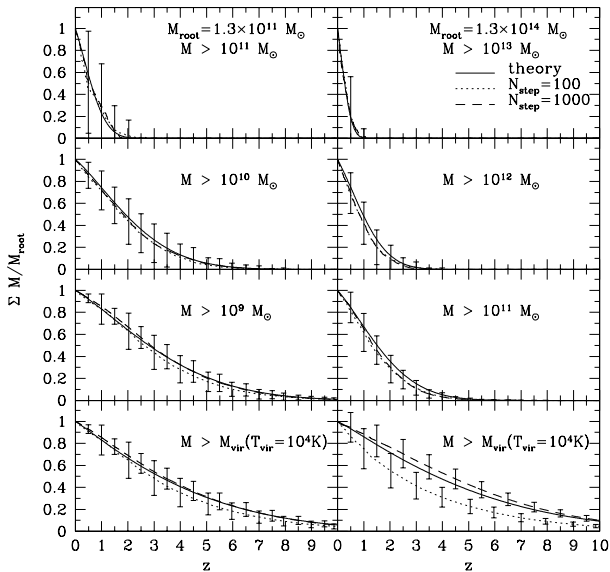}
\caption{Cumulative mass fraction of progenitors of mass exceeding a
given threshold, $M_{\rm thre}$, for $M_{\mathrm{root}}=1.3\times
10^{11}\;M_{\odot}$ (\textit{Left}) and $1.3\times 10^{14}\;M_{\odot}$
(\textit{Right}). The dotted and dashed curves indicate the averages
from $N_{\rm step}=100$ and 1000 merger tree realizations with the
corresponding Poisson error-bars.  \label{fig:massfrac}}
\end{figure}
%%%%%%%%%%%%%%%%%%%%%%%%%%%%%%%%%%%%%%%%%%%%%%%%%%%%%%%%%%%%%%%%%%%%%%%%%

The difference between $N_{\mathrm{step}}=100$ and 1000 is more
clearly illustrated when we plot the cumulative mass fraction of
progenitors of mass exceeding a threshold value of $M_{\rm thre}$.  More
specifically, figure~\ref{fig:massfrac} compares the theoretical
prediction,
%%%%%%%%%%%%%%%%%%%%%%%%%%%%%%%%%%%%%%%%%%%%%%%%%%%%%%%%
\begin{equation}
 F_{\rm th}(>M_{\mathrm{thre}}; z) =
 \int^{M_{\mathrm{root}}}_{M_{\rm thre}}\!\!\!\!dM
 \frac{dP}{dM}(M,z|M_{\mathrm{root}},z_{\mathrm{min}}),
 \label{eq:fracpth}
\end{equation}
%%%%%%%%%%%%%%%%%%%%%%%%%%%%%%%%%%%%%%%%%%%%%%%%%%%%%%%%
with the average from the tree realizations,
%%%%%%%%%%%%%%%%%%%%%%%%%%%%%%%%%%%%%%%%%%%%%%%%%%%%%%%%
\begin{equation}
 F_{\rm model}(>M_{\rm thre}; z)
 =\frac{1}{N_{\mathrm{ens}}M_{\mathrm{root}}}
 \sum_{M_{\mathrm{halo}}(z)\ge
 M_{\rm thre}}\mbox{\hspace{-7mm}}M_{\rm halo}(z),
\end{equation}
%%%%%%%%%%%%%%%%%%%%%%%%%%%%%%%%%%%%%%%%%%%%%%%%%%%%%%%%
for $M_{\rm root} = 1.3\times10^{11}\;M_{\odot}$ (\textit{Left}) and
$1.3\times10^{14}\;M_{\odot}$ (\textit{Right}).  In all panels shown here,
the results with $N_{\rm step}=1000$ better reproduce the theoretical
prediction, mainly because of the small-scale behavior (see
figures~\ref{fig:100Step} and \ref{fig:1000Step}).  Since these small
mass progenitors at earlier redshifts significantly contribute to
radiative cooling and thereby subsequent star formation in the entire
halo, this difference is indeed critical in the Monte-Carlo modeling
of galaxy formation. We discuss the effect on gas cooling
explicitly in the next section.

%%%%%%%%%%%%%%%%%%%%%%%%%%%%%%%%%%%%%%%%%%%%%%%%%%%%%%%%%%%%%%%%%%%%%%%%
\begin{figure}[thb]
\centering \FigureFile(60mm,60mm){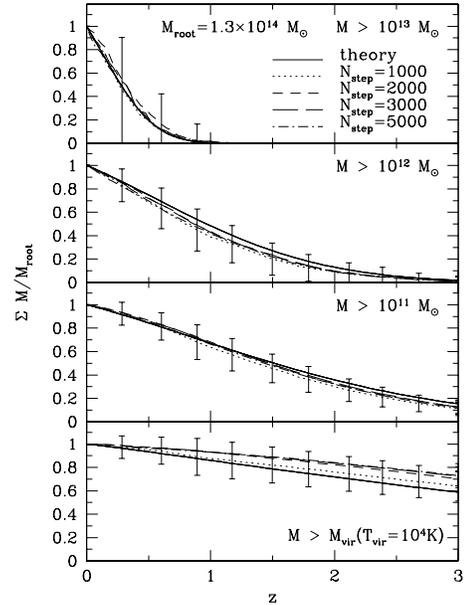}
  \caption{Cumulative mass fraction of progenitors with different values
    of $N_\mathrm{step}$ for $M_{\mathrm{root}}=1.3\times
    10^{14}\;M_{\odot}$.}
  \label{fig:mflargeNstep}
\end{figure}
%%%%%%%%%%%%%%%%%%%%%%%%%%%%%%%%%%%%%%%%%%%%%%%%%%%%%%%%%%%%%%%%%%%%%%%%

We interpret the above as an empirical result due to the balance
between the mass-weighted probability and the timesteps. If the use of
the mass-weighted probability \textit{were} strictly justified, the
proper realizations with a smaller timestep would become more difficult
numerically, and there is no reason why we could obtain better
agreement with a larger $N_{\rm step}$. On the other hand, we understand
that the mass-weighted probability is nothing but a phenomenological
remedy of the problem, and with this choice $N_{\rm step}=1000$ seems
to work better than $N_{\rm step}=100$ empirically.

In fact, still larger values of $N_{\rm step}$ do not necessarily
improve the result.  Figure~\ref{fig:mflargeNstep} is a similar plot
as figure~\ref{fig:massfrac} for $M_{\rm
  root}=1.3\times10^{14}\;M_\odot$, but with increasing
$N_\mathrm{step}$. The cumulative mass fraction for
$M>10^{11}\;M_\odot$ is almost unchanged, but the contribution from
smaller mass progenitors steadily increases as $N_\mathrm{step}$
becomes larger. This reflects the fact that the empirical use of the
mass-weighted conditional probability does not guarantee 
convergence of the result with respect to $N_\mathrm{step}$.

We thus conclude that $N_{\mathrm{step}}\sim 1000$ is the optimal value
to reproduce the Press--Schechter mass function in our method.

\subsection{Number of Progenitors}

\begin{figure}[thb]
  \centering \FigureFile(60mm,60mm){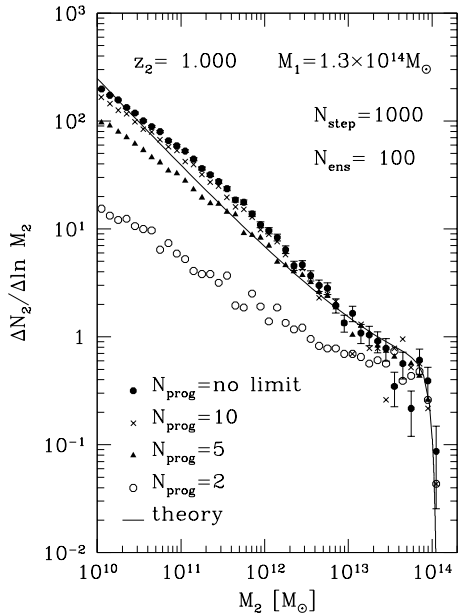}
\caption{Effect of the upper limit on the number of progenitors,
$N_{\rm prog}$, on the progenitor distribution function for
$M_{\mathrm{root}}=1.3\times 10^{14}\;M_{\odot}$ at $z=1.0$ from $N_{\rm
ens}=100$ merger tree realizations with $N_{\rm step}=1000$.  The solid
line is the number-weighted probability function,
equation~(\ref{eq:eps-num}). The filled circles indicate the results with all
progenitors satisfying the mass conservation
[equation~(\ref{eq:massconserve})], while crosses, filled triangles, and open
circles are for $N_{\rm prog} = 10$, 5, and 2, respectively.
\label{fig:Nprog}}
\end{figure}

Finally, we briefly discuss how many progenitors, $N_{\rm prog}$, one
should keep in order to properly reproduce the merger trees.
Figure~\ref{fig:Nprog} displays the distribution functions at $z=1$
for the merger tree of $M_{\mathrm{root}}=1.3\times 10^{14}\;M_{\odot}$ at
$z=0$. In this particular example, we use $N_{\rm step}=1000$ and the
results are averaged over $N_{\rm ens}=100$ realizations for each
$N_{\rm prog}$. Obviously a smaller value of $N_{\rm prog}$ does not
properly link the merger tree back to higher redshifts, and the number
of small-mass halos is systematically under-predicted compared with
the extended Press--Schechter model (solid curve).  While we do not set
any upper limit on $N_{\mathrm{prog}}$, figure~\ref{fig:Nprog}
indicates that $N_{\mathrm{prog}} \gtrsim 5$ is acceptable given the
accuracy of the present scheme.  Although some authors employ a binary
merger tree in Monte-Carlo modeling, that scheme needs to be
adjusted with a careful choice of the timestep and other parameters.

%%%%%%%%%%%%%%%%%%%%%%%%%%%%%%%%%%%%%%%%%%%%%%%%%%%%%%%%%%%%%%%%%%%%%%%%
%%%%%%%%%%%%%%%%%%%%%%%%%%%%%%%%%%%%%%%%%%%%%%%%%%%%%%%%%%%%%%%%%%%%%%%%

In conclusion, we have found that a reasonable agreement between the theory
and the merger tree realizations can be obtained by employing the
mass-weighted conditional probability into the Somerville and
Kolatt~(1999) scheme 
with $N_{\mathrm{step}}\sim 1000$.

\section{Gas Cooling}

So far, we have restricted our discussion to the gravitational aspect
of the halo evolution.  We next consider how the resulting merger
trees affect the cold gas fraction in an individual halo. Of course, we
must eventually discuss the effects on the efficiency of star
formation, but we focus on gas cooling alone, since modeling star
formation, the feedback from supernovae, the chemical evolution and so
on necessarily introduce additional (numerical) parameters and the
interpretation becomes more complicated.  Thus, the principal aim of
this section is to see if we can achieve a convergence of the cold gas
fraction from different realizations of the merger trees.

\subsection{Description of Gas Cooling}

Our prescription of gas cooling in the merger trees goes as follows.
First, we assume that the density profile of dark halos obeys the
universal shape (\cite{nfw96}):
%%%%%%%%%%%%%%%%%%%%%%%%%%%%%%%%%%%%%%%%%%%%%%%%%%%%%%%%%%%%%%%%%%%%%%%%%
\begin{equation}
\label{eq:nfw}
 \rho_{\rm halo}(r;M)=
 \left\{\begin{array}{cc}
  \displaystyle
  \frac{\bar{\rho}(z) \, \delta_\mathrm{c}}{
    (r/r_\mathrm{s}) (1+r/r_\mathrm{s})^2} & (r< r_{\rm vir}) \\
  \displaystyle
  0 & (r>r_{\rm vir}),
 \end{array}\right. 
\end{equation}
%%%%%%%%%%%%%%%%%%%%%%%%%%%%%%%%%%%%%%%%%%%%%%%%%%%%%%%%%%%%%%%%%%%%%%%%%
where $\bar\rho(z) \equiv \Omega_0 \rho_{\rm c0} (1+z)^3$ is the mean
density of the universe at $z$, $\rho_{\rm c0}$ is the present
critical density, $\delta_\mathrm{c}(M)$ is the characteristic density
excess, and $r_{\rm vir}(M)$ and $r_\mathrm{s}(M)$ indicate the virial
radius and the scale radius of the halo, respectively.

The virial radius is defined according to the spherical collapse model
as
%%%%%%%%%%%%%%%%%%%%%%%%%%%%%%%%%%%%%%%%%%%%%%%%%%%%%%%%%%%%%%%%%%%%%%%%%
\begin{equation}
 r_{\rm vir}(M) \equiv
 \left(\frac{3M}{4\pi\bar{\rho} \Delta_{\rm nl}}\right)^{1/3} ,
\label{eq: r_vir}
\end{equation}
%%%%%%%%%%%%%%%%%%%%%%%%%%%%%%%%%%%%%%%%%%%%%%%%%%%%%%%%%%%%%%%%%%%%%%%%%
and useful approximation for the critical over-density,
$\Delta_{\rm nl}= \Delta_{\rm nl}(\Omega_0,\lambda_0)$, may be found in
Kitayama and Suto (1996). The two parameters, $r_{\rm s}$ and $r_{\rm
  vir}$, are related in terms of the concentration parameter,
%%%%%%%%%%%%%%%%%%%%%%%%%%%%%%%%%%%%%%%%%%%%%%%%%%%%%%%%%%%%%%%%%%%%%%%%%
\begin{equation}
  \label{eq: concentration}
c(M,z) \equiv \frac{r_{\rm vir}(M,z)}{r_\mathrm{s}(M,z)}.    
\end{equation}
%%%%%%%%%%%%%%%%%%%%%%%%%%%%%%%%%%%%%%%%%%%%%%%%%%%%%%%%%%%%%%%%%%%%%%%%%
We use an approximate fitting function from the simulation data of
Bullock et al.\ (2001),
%%%%%%%%%%%%%%%%%%%%%%%%%%%%%%%%%%%%%%%%%%%%%%%%%%%%%%%%%%%%%%%%%%%%%%%%%
\begin{equation}
 c(M,z)=\frac{8.0}{1+z}\;\left(\frac{M}{10^{14}\;M_{\odot}}\right)^{-0.13}. 
\label{eq: c_Bullock}
\end{equation}
%%%%%%%%%%%%%%%%%%%%%%%%%%%%%%%%%%%%%%%%%%%%%%%%%%%%%%%%%%%%%%%%%%%%%%%%%
The condition that the total mass inside $r_{\rm vir}$ be equal to $M$
relates $\delta_{\rm c}$ to $c$ as
%%%%%%%%%%%%%%%%%%%%%%%%%%%%%%%%%%%%%%%%%%%%%%%%%%%%%%%%%%%%%%%%%%
\begin{equation}
  \delta_\mathrm{c} = {\Delta_{\rm nl} \over 3} {c^3 \over \ln(1+c)
  -c/(1+c)}. 
\end{equation}
%%%%%%%%%%%%%%%%%%%%%%%%%%%%%%%%%%%%%%%%%%%%%%%%%%%%%%%%%%%%%%%%%%

\citet{MSS97} showed that if the hot gas is isothermal and in
hydrostatic equilibrium, the gas density profile is well approximated
by the isothermal $\beta$-model,
%%%%%%%%%%%%%%%%%%%%%%%%%%%%%%%%%%%%%%%%%%%%%%%%%%%%%%%%%%%%%%%%%%%
\begin{equation}
 \rho_{\rm hot}(r) =
 \frac{\rho_{\rm hot, 0}}{[1+(r/r_\mathrm{c})^{2}]^{3\beta/2}},
 \label{eq:rhogas}
\end{equation}
%%%%%%%%%%%%%%%%%%%%%%%%%%%%%%%%%%%%%%%%%%%%%%%%%%%%%%%%%%%%%%%%%%%
where $r_\mathrm{c}\sim 0.22r_\mathrm{s}$. We fix $\beta=2/3$ for
simplicity.  The amplitude $\rho_{\rm hot, 0}$ is computed so as to
reproduce the total hot gas in the halo when integrated up to
$r=r_{\rm vir}$.  The hot gas is gradually converted to cold gas
according to the prescription below, but still the total baryon (hot +
cold) fraction within the virial radius of each halo is set to the
cosmic average, $\Omega_{\rm B}/\Omega_0$.

Once the gas profile is specified, one can compute the cooling
timescale at radius $r$ from the center of the halo,
%%%%%%%%%%%%%%%%%%%%%%%%%%%%%%%%%%%%%%%%%%%%%%%%%%%%%%%%%%%%%%%%%%%
\begin{equation}
\label{eq:tcool}
 t_{\mathrm{cool}}=\frac{3}{2}
 \frac{\rho_{\mathrm{hot}}(r)}{\mu m_{\mathrm{p}}}
 \frac{k_{\mathrm{B}}T_{\mathrm{gas}}}
 {\Lambda(T_{\mathrm{gas}})n_{\mathrm{H}}^{2}(r)}, 
\end{equation}
%%%%%%%%%%%%%%%%%%%%%%%%%%%%%%%%%%%%%%%%%%%%%%%%%%%%%%%%%%%%%%%%%%%
where $\mu$ is the mean molecular weight, $m_{\mathrm{p}}$ the
proton mass, $k_{\mathrm{B}}$ the Boltzmann constant,
$\Lambda(T_{\mathrm{gas}})$ the radiative cooling function for gas of
temperature $T_{\mathrm{\mathrm{gas}}}$, and $n_{\mathrm{H}}(r)$ the 
number density of hydrogen (including both neutral and ionized).  We
assume that the gas has the primordial abundance of hydrogen and
helium ($X=0.76$ and $Y=0.24$, and thus $n_{\rm H}=\rho_{\rm
  hot}/Xm_{\rm p}$), and compute the corresponding cooling function
$\Lambda(T)$ (e.g., Sasaki and Takahara 1994). Since we neglect the
molecular and metal cooling, $\Lambda(T)=0$ at $T\le
10^{4}\;\mathrm{K}$.

%%%%%%%%%%%%%%%%%%%%%%%%%%%%%%%%%%%%%%%%%%%%%%%%%%%%%%%%%%%%%%%%%%%%%%%%%
\begin{figure}[thb]
  \centering \FigureFile(60mm,60mm){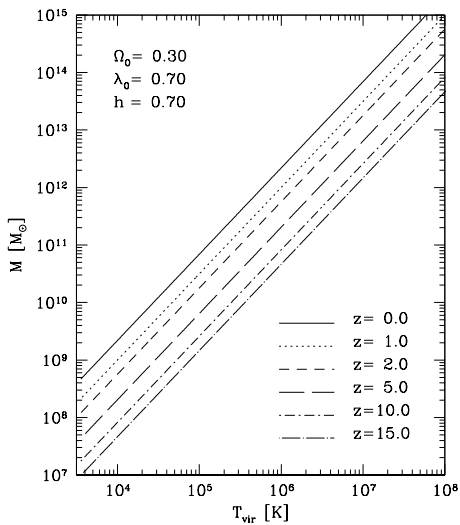}
\caption{Mass of halos as a function of the virial temperature at
  different redshifts in the $\Lambda$CDM model.  \label{fig:tvirtom}}
\end{figure}
%%%%%%%%%%%%%%%%%%%%%%%%%%%%%%%%%%%%%%%%%%%%%%%%%%%%%%%%%%%%%%%%%%%%%%%%%

We further assume that the temperature of the hot gas,
$T_{\mathrm{gas}}$, is equal to the virial temperature,
$T_{\mathrm{vir}}$, of the halo.  The relation between the virial
temperature and the mass of a halo is plotted in
figure~\ref{fig:tvirtom}. We can then solve equation (\ref{eq:tcool})
for the cooling radius, $r_{\mathrm{cool}}$, within which the gas can
cool within a given cooling timescale ($\tau_{\rm cool}$):
%%%%%%%%%%%%%%%%%%%%%%%%%%%%%%%%%%%%%%%%%%%%%%%%%%%%%%%%%%%%%%%%%%%%%%%
\begin{equation}
\hspace*{-0.5cm} 
r_{\rm cool}(\tau_{\rm cool}) \equiv 0.22 \frac{r_{\rm vir}(M,z)}{c(M,z)}
 \sqrt{\frac{2\mu \Lambda(T_{\mathrm{vir}})
 \rho_{\rm hot,0}\tau_{\rm cool}} {3m_\mathrm{p} k_{\rm B}T_{\rm vir}}
 - 1 } . 
\end{equation}
%%%%%%%%%%%%%%%%%%%%%%%%%%%%%%%%%%%%%%%%%%%%%%%%%%%%%%%%%%%%%%%%%%%%%%%
It now remains to define the origin of $\tau_{\rm cool}$. Actually,
this is fairly arbitrary in a sense, and we adopt the following simple
picture. When a halo of mass $M_{\rm f}$ \textit{forms} at the
formation redshift, $z_{\rm f}$, its hot gas is supposed to reach the
profile [equation~(\ref{eq:rhogas})] instantaneously. This is defined to be
the origin of $\tau_{\rm cool}$ for the halo. In the subsequent
timesteps, we neglect the change in the hot gas profile even if the
halo mass ($M$) grows due to mergers and the cooling radius is computed
with $\tau_{\rm cool}$ set to the elapsed cosmic time since $z_{\rm
  f}$. When $M$ exceeds $2M_{\rm f}$, the halo is replaced by a
\textit{newly} formed massive halo, and the hot gas profile is reset
to the profile [equation~(\ref{eq:rhogas})] corresponding to the new mass
and the virial temperature, and we reset the origin of $\tau_{\rm
  cool}$ as the new formation epoch. Incidentally, we made sure that
the value of $1.5M_{\rm f}$ instead of $2M_{\rm f}$ does not change
the result, which is consistent with the finding of Cole et al.\
(2000).  We apply this procedure for all halos, and the cold gas in
each progenitor halo is simply accumulated (without reheated)
according to the merger trees.

\subsection{Cold Gas Fraction in the Monte-Carlo Realization of
  Merging Histories}

In a practical implementation of the merger tree algorithm, one has to
stop tracing the progenitors of halos of mass below the resolution
mass, $M_{\rm res}$. We discuss the relevance of our choice of $M_{\rm
  res}$ by looking at the cold gas fraction.  Previous authors often
apply the cutoff at a fixed mass or a circular velocity of halos; for
instance, \citet{Cole00} consider halos with $M > 5\times
10^{9}h^{-1}\;M_{\odot}$ in their merger trees, while Somerville and
Primack~(1999) take 
account of halos with the circular velocity exceeding
$40\;\mathrm{km\ s}^{-1}$ (corresponding to the virial temperature $T_{\rm
  vir}\sim 6\times10^{4}\;\mathrm{K}$).  The latter condition comes
from an estimate of the smallest scale of halos which can cool in the
presence of the UV background  (e.g., \cite{TW96}; \cite{KI00}).

In our present analysis, the cooling function, $\Lambda(T)$, vanishes
below $T=10^4\;\mathrm{K}$, since we neglect both the UV heating and
the metal/molecular cooling. Thus, we set the resolution mass, $M_{\rm
  res}(z)$, as $M(T_{\rm vir}=10^4\;{\rm K})$.  As
figure~\ref{fig:tvirtom} shows, this scale increases rapidly with time,
resulting in a significant improvement of the computing time. On
the other hand, this might systematically underestimate the cold gas
fraction, since halos of mass below $M_{\rm res}(z_1)$ may have
progenitors of mass larger than $M_{\rm res}(z_2)$ at the earlier
epoch ($z_2>z_1$).  Fortunately, this is not an important effect, as we
show below.

%%%%%%%%%%%%%%%%%%%%%%%%%%%%%%%%%%%%%%%%%%%%%%%%%%%%%%%%%%%%%%%%%%%%%%%%%
\begin{figure}[thb]
  \centering \FigureFile(60mm,60mm){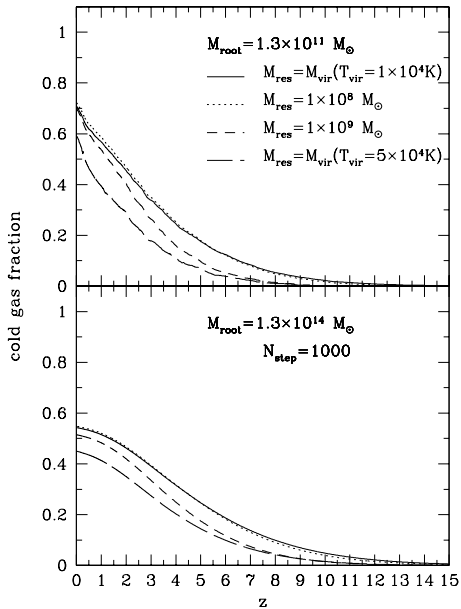}
  \caption{Cold gas fraction [equation~(\ref{eq:coldgasfrac})] with different
    resolution mass, $M_{\rm res}$, averaged from $N_{\rm ens}=10$
    merging tree realizations at $N_{\rm step}=1000$. Upper and lower
    panels show the results for $M_{\rm root}= 1.3\times 10^{11}\;M_{\odot}$
    and $1.3\times 10^{14}\;M_{\odot}$; $M_{\rm res}(z)=M(T_{\rm
      vir}=10^4\;{\rm K})$, $M(T_{\rm vir}=5\times10^4\;{\rm K})$,
    $10^8\;M_{\odot}$, and $10^9\;M_{\odot}$.
    \label{fig:convmres}}
\end{figure}
%%%%%%%%%%%%%%%%%%%%%%%%%%%%%%%%%%%%%%%%%%%%%%%%%%%%%%%%%%%%%%%%%%%%%%%%%

To see this in detail, we plot in figure~\ref{fig:convmres} the cold
gas fraction averaged over all progenitors at $z$ of a root halo of
mass $M_{\rm root}$ at $z=0$,
%%%%%%%%%%%%%%%%%%%%%%%%%%%%%%%%%%%%%%%%%%%%%%%%%%%%%%%%%%%%%%%%%%%%%%%
\begin{eqnarray}
\label{eq:coldgasfrac}
f_{\rm cold}(z;M_{\rm root}) \equiv 
\frac{\Omega_0}{\Omega_{\rm B}M_{\rm root}} 
\sum_{M_{\rm prog}>M_{\rm res}} M_{\rm cold}(M_{\rm prog}) ,
\end{eqnarray}
%%%%%%%%%%%%%%%%%%%%%%%%%%%%%%%%%%%%%%%%%%%%%%%%%%%%%%%%%%%%%%%%%%%%%%%
for a merger tree with $N_{\rm step}=1000$.  If we adopt a constant
value for the resolution mass, $M_{\rm res} < 10^8\;M_{\odot}$ yields the
convergent result for the cold gas fraction. Exactly the same
convergence is obtained for the time-dependent $M_{\rm res}(z)$ when
the value is set to $M(T_{\rm vir}=10^4\;{\rm K})$, but not if we use
$M(T_{\rm vir}=5\times 10^4\;{\rm K})$, for instance.  This critical
value is expected to vary depending on the thermal history of the
universe, but the appropriate value for $T_{\rm vir}$ is
straightforwardly read off from the relevant cooling function.
Actually, $M_{\rm res}(z)$ increases in this case and exceeds
$10^9\;M_{\odot}$ (see figure~\ref{fig:tvirtom}). Thus, the required merging
tree is less demanding from a computational point of view than that
for $M_{\rm res}=10^{8}\;M_{\odot}$, for instance, as illustrated in
table~\ref{tab:cputime}.  Thus we decide to choose $M_{\rm res}(z) =
M(T_{\rm vir}=10^4\;{\rm K})$ for gas with the primordial abundance.

\begin{figure}[thb]
  \centering \FigureFile(60mm,60mm){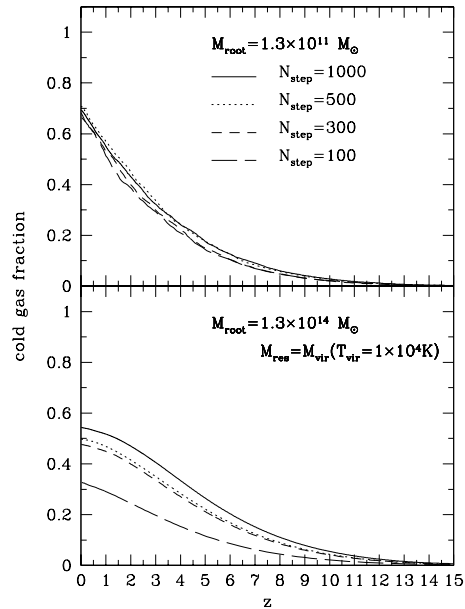}
  \caption{Cold gas fraction averaged from $N_{\rm ens}=10$ merger tree
    realizations with different $N_{\rm step}$.  The upper and lower
    panels show the results for $M_{\rm root}= 1.3\times 10^{11}\;M_{\odot}$
    and $1.3\times 10^{14}\;M_{\odot}$.  The resolution mass of the tree is
    fixed as $M_{\rm res}(z)=M(T_{\rm vir}=10^4\;{\rm K})$.
  \label{fig:convnsamp}}
\end{figure}

Finally, we show the convergence with respect to $N_{\rm step}$.
Figure~\ref{fig:convnsamp} plots the cold gas fraction for $M_{\rm
  root}=1.3\times 10^{11}\;M_{\odot}$ (\textit{Upper}) and $1.3\times
10^{14}\;M_{\odot}$ (\textit{Lower}) for merging trees with different
$N_{\rm step}$.  The results are fairly in agreement for small $M_{\rm
  root}$, but are very different for large $M_{\rm root}$. This is
because the merger tree at small mass scales, especially at $M(T_{\rm
  vir}=10^4 {\rm K}) < M < 10^{10}M_\odot$, is well reproduced only
when we use $N_{\rm step}=1000$ (see figure~\ref{fig:massfrac}).  When
we repeat the same calculation sampling every 10 steps from the
$N_{\rm step}=1000$ tree, the result is almost indistinguishable.
Thus, we conclude that the progenitor distribution at small scales is
quite essential in the estimate of the cold mass fraction of large
halos.

Incidentally the use of the timestep much smaller than $t_{\rm dyn,
  vir}(z)$ enables one to describe the collapse and gas cooling more
realistically than the instantaneous approximation. While we do not
attempt this in the present paper, this would improve the estimate of
the cold gas fraction quantitatively.

%%%%%%%%%%%%%%%%%%%%%%%%%%%%%%%%%%%%%%%%%%%%%%%%%%%%%%%%%%%%%%%%%%%%%%%%%
%%%%%%%%%%%%%%%%%%%%%%%%%%%%%%%%%%%%%%%%%%%%%%%%%%%%%%%%%%%%%%%%%%%%%%%%%

\section{Conclusions and Discussion}

We attempted several convergence tests of the merger trees generated
with the Monte-Carlo method. While this method provides a useful tool
for modeling galaxy formation in a complementary manner to more
intensive cosmological simulations with ad hoc recipes of galaxy
formation (e.g., Cen, Ostriker 1992; Weinberg et al.\ 1997; Yoshikawa
et al.\ 2001), the lack of an explicit expression for the joint
distribution function of progenitors [equation~(\ref{eq:jointprob})]
requires one to put an additional assumption in practice. We confirmed
that a repeated use of the \textit{mass-weighted} conditional
probability [equation~(\ref{eq:eps-num})] reasonably reproduces the
progenitor distribution predicted in the extended Press--Schechter
theory if one adopts fairly small timesteps in redshift, $N_{\rm step}
\sim 1000$, a factor of ten larger than a typical value used in
previous work. We note, however, that one can alternatively achieve
a similar result by fine-tuning the timestep as a function of
$M_{1}$ (e.g., \cite{SK99}) instead of equation~(\ref{eq:zbin}), as we
adopted here.

One may avoid the above problem also by using merger trees generated
via $N$-body simulations (\cite{gif99a}; Somerville et al.\ 2001). In
fact, they claim that the agreement between the $N$-body simulations and
the Monte-Carlo method is good. Benson et al.\ (2001) compared the SPH
simulations and the Monte-Carlo modeling, and concluded that both
agree with each other on the cold gas mass fraction and mass function
of the halos.  While this comparison is encouraging, it is not yet
clear if the lack of the joint distribution function of progenitors
[equation~(\ref{eq:jointprob})] in the Monte-Carlo modeling may not be
essential. Thus, further detailed studies are definitely important to
test the reliability of {\it both\/} $N$-body and the Monte-Carlo
modeling in generating merger tree realizations.

\vspace*{0.5cm}

We thank Kazuhiro Shimasaku and Tomonori Totani for discussions and
suggestions in the early phase of this work.  This research was
supported in part by the Grant-in-Aid from Monbu-Kagakusho, Japan
(07CE2002, 12304009, 12640231).  T.K. gratefully acknowledges support
from Research Fellowships of the Japan Society for the Promotion of
Science for Young Scientists (7202).

\onecolumn

%%%%%%%%%%%%%%%%%%%%%%%%%%%%%%%%%%%%%%%%%%%%%%%%%%%%%%%%%%%%%%%%%%%%%%%%

%%%%%%%%%%%%%%%%%%%%%%%%%%%%%%%%%%%%%%%%%%%%%%%%%%%%%%%%%%%%%%%%%%%%%%%%

%\onecolumn

\vspace*{1cm}

%%%%%%%%%%%%%%%%%%%%%%%%%%%%%%%%%%%%%%%%%%%%%%%%%%%%%%%%%%%%%%%%%%%%%%%%%
\begin{table}[thb]
  \caption{Summary of the variables used in building merger trees
    and in gas cooling.  \label{tab:parameters}}
  \begin{center}
    \begin{tabular}{ccl}
      \hline \hline
      Symbol & Adopted value & Physical meaning \\ \hline
      $M_{\rm root}$ & --- & mass of halo at $z=z_{\mathrm{min}}$ \\
      $T_{\rm vir}$ & --- & virial temperature of halo \\
      $M_{\rm res}$ & $M(T_{\rm vir}=10^4\;{\rm K})$ & minimal mass of
      progenitors resolved in each merger tree \\
      $\tau_{\rm cool}$ & halo mass doubling time & cooling time scale
      for gas in hosting halos \\
      $N_{\rm step}$ & 1000 
      & number of  redshift bins (logarithmically equal interval)\\
      $N_{\rm ens}$ & --- & number of realizations of merger trees\\
      $z_{\rm min}$ & 0 & minimum redshift of merger trees \\
      $z_{\rm max}$ & 15 & maximum redshift of  merger trees \\ \hline
    \end{tabular}
  \end{center}
\end{table}
%%%%%%%%%%%%%%%%%%%%%%%%%%%%%%%%%%%%%%%%%%%%%%%%%%%%%%%%%%%%%%%%%%%%%%%%%

%%%%%%%%%%%%%%%%%%%%%%%%%%%%%%%%%%%%%%%%%%%%%%%%%%%%%%%%%%%%%%%%%%%%%%%%%
\begin{table}[thb]
  \caption{CPU timing of the Monte-Carlo modeling for one merger tree on
    a 21264 alpha $600\;\mathrm{MHz}$ machine.
    \label{tab:cputime}}
  \begin{center}
    \begin{tabular}{cccrrr}\hline \hline
      \multicolumn{1}{c}{$M_\mathrm{root}$}&
      \multicolumn{1}{c}{$N_\mathrm{step}$}&
      \multicolumn{1}{c}{$M_\mathrm{res}$ }&
      \multicolumn{1}{r}{Number of progenitors}&
      \multicolumn{2}{c}{CPU-time (s)}\\
      \multicolumn{1}{c}{$(M_{\odot})$}&
      \multicolumn{1}{c}{}&
      \multicolumn{1}{c}{}&
      \multicolumn{1}{r}{}&
      \multicolumn{1}{r}{merger tree}&
      \multicolumn{1}{r}{gas cooling}\\\hline 
      $1.3\times 10^{11}$ & 100 & $10^8\;M_{\odot}$ 
      &3876& 3.4 & 0.55\\
      $1.3\times 10^{11}$ & 100 & $M(T_\mathrm{vir}=10^4\;\mathrm{K})$
      &1687& 2.9 & 0.43\\
      $1.3\times 10^{11}$ & 1000 & $10^8\;M_{\odot}$ 
      &56057& 15.1 & 4.4\\
      $1.3\times 10^{11}$ & 1000 & $M(T_\mathrm{vir}=10^4\;\mathrm{K})$
      &20556& 6.8 & 2.3\\\hline
      $1.3\times 10^{14}$ & 100 & $10^8\;M_{\odot}$ 
      &1122420& 244 & 75.6\\
      $1.3\times 10^{14}$ & 100 & $M(T_\mathrm{vir}=10^4\;\mathrm{K})$
      &640409& 137 & 46.5\\
      $1.3\times 10^{14}$ & 1000 & $10^8\;M_{\odot}$ 
      &29725672& 6514 & 1993\\
      $1.3\times 10^{14}$ & 1000 & $M(T_\mathrm{vir}=10^4\;\mathrm{K})$
      &18875982& 4140 & 1400\\ \hline
    \end{tabular}
  \end{center}
\end{table}
%%%%%%%%%%%%%%%%%%%%%%%%%%%%%%%%%%%%%%%%%%%%%%%%%%%%%%%%%%%%%%%%%%%%%%%%%

\end{document}